# Painting with Hue, Saturation, and Brightness Control by Nanoscale 3D Printing

Hao Wang,[1] Qifeng Ruan,[1] Hongtao Wang,[1] Soroosh Daqiqeh Rezaei,[1] Kevin T. P. Lim,[2] Hailong Liu,[1] Wang Zhang,[1] Jonathan Trisno,[1] John You En Chan,[1] Joel K.W. Yang[1, 3, *]

[1]*Engineering Product Development Pillar, Singapore University of Technology and Design, 8 Somapah Road, Singapore 487372, Singapore*

[2]*Cavendish Laboratory, JJ Thomson Ave, Cambridge, CB3 0HE, United Kingdom*

[3]*Institute of Materials Research and Engineering, A\*STAR (Agency for Science, Technology and Research), 2 Fusionopolis Way, #08-03 Innovis, 138634 Singapore*

[*]joel_yang@sutd.edu.sg

**Abstract**

Varying only the in-plane or out-of-plane dimensions of nanostructures produces a wide range of colourful elements in metasurfaces and thin films. However, achieving shades of grey and control of colour saturation remains challenging. Here, we introduce a hybrid approach to colour generation based on the tuning of nanostructure geometry in all three dimensions. Through two-photon polymerization lithography, we systematically investigated colour generation from the simple single nanopillar geometry made of low-refractive-index material; realizing grayscale and full colour palettes with control of hue, saturation, brightness through tuning of height, diameter, and periodicity of nanopillars. Arbitrary colourful and grayscale images were painted by mapping desired prints to precisely controllable parameters during 3D printing. We extend our understanding of the scattering properties of the low-refractive-index nanopillar to demonstrate grayscale inversion and colour desaturation, with steganography at the level of single nanopillars.

Structural colours arise from the interaction of light with nanostructured materials that are typically colourless in bulk. Compared to dyes, structural colours are potentially more robust against photobleaching and can achieve ultra-high resolution colour prints with nanoscale precision[1-6]. New mechanisms for structural colour generation have been discovered in recent years. Uniquely, individual nanostructures in metals and high-refractive-index dielectrics generate colours through plasmonics and Mie resonances, without varying its chemical composition or requiring many repeating elements[7-17]. Conversely, low-refractive-index materials generally require larger dimensions to generate colours, e.g., periodic structures such as photonic crystals, microdroplets, and thin films with characteristic length scales of multiple wavelengths, thus reducing the level of control and resolution on the colours produced. As low-refractive-index materials such as plastics and biodegradable polymers are ubiquitous, identifying new mechanisms to generate colours with subwavelength dimensions with these transparent and cost-effective materials could enable the design of dye-free colourful devices with unique decorative and spectral properties, while being recyclable and compatible with the circular economy[18].

Compared to the red, green, and blue (RGB) model, the hue, saturation, and brightness (HSB) color space is a closer representation of the way in which humans perceive colour. In addition to different hues, different shades of grey and saturation levels are also needed to achieve arbitrary color prints. However, an elegant solution to control the HSB channels in structural colours is still elusive. Prior work is based on the subtractive colour mixing method, i.e., by combining multiple nano-antennae of different geometries/sizes into a super pixel to obtain grayscale[19-23]. This method requires complex colour mixing design and test, and also increases the size of a single pixel. Furthermore, larger sizes of pixels are necessary if more levels of grey are needed. Different grey levels are achieved by a weighted adjustment of the primary colours, hence reducing print resolution. Random nanostructured films can also be used to control grey levels. However, high-resolution printing with multilevel grayscale and vivid colours is limited by random scattering of light in these disordered structures[24].

Due to restrictions in conventional nanofabrication approaches, the tunability of colour elements is either constrained to in-plane (e.g., for plasmonic and high-refractive-index dielectric metasurfaces), or out-of-plane parameters in thin films (e.g., thin absorptive films and Fabry–Pérot cavities), but rarely both[25,26]. Out-of-plane control is challenging in general, as it requires either multiple deposition procedures or precise grayscale lithography within a limited range of thickness[27-31]. Consequently, one necessarily increases design complexity to achieve a full range of colours, e.g., through the incorporation of multiple materials[7], interleaved sub-pixels[20] or sacrificing polarization independence[21]. As the HSB color space is three dimensional (3D) in nature, we hypothesize that full control of HSB could be achieved elegantly while maintaining nanoscale pixel sizes if one could control nanostructure dimensions for both in-plane ($x$, $y$) and out-of-plane ($z$) directions.

Nanoscale additive manufacturing lifts the constraints of conventional binary lithography and opens a full 3D nanostructure design space. Advantages include lower capital investment, design customization, waste minimization, and rapid prototyping[32]. In particular, two-photon polymerization lithography (TPL) makes it possible to print structures with hundred-nanometer length scales[33], thus making it applicable in visible-light optics to produce colourful photonic crystal structures[26,34,35], high-quality holograms free of the zero-order spot[36], holographic colour prints[37], moiré multiplexed information retrieval with microlenses[38], vortex beam[39], circular polarizer[40], chiral beamsplitter[41], ultracompact multi-lens system on fibre[42], achromatic metalenses[43], and hybrid photonic integrated chip[44].

Here, we realize the first full colour palette in 3D color space, using a simple design approach that controls colours at the single nanopillar level. The three degrees of freedom of the 3D full colour palette correspond to the three printing parameters of the nanopillar, i.e., height (H), exposure time (T), and period (P). Different color hues, saturation, and brightness are achieved, enabling us to produce colourful and grayscale images by splitting them into H, T, and P channels, or HTP in short. The optical response of a nanopillar is analyzed from multipole decomposition, with good agreement between simulated spectra and experiments. Furthermore, grayscale inversion and colour desaturation are investigated with brightfield and darkfield illumination, and a reliable height-based steganography method with single nanopillars is demonstrated. Our results highlight the potential in realizing a 3D colour design space using a simple geometry and with a single patterning step by direct laser writing. As the refractive index of the photoresists is comparable to the materials used in nature to generate structural colours, e.g., chitin and cellulose, this approach could be useful in the prototyping of 3D models or devices made of complex bio-inspired structures.

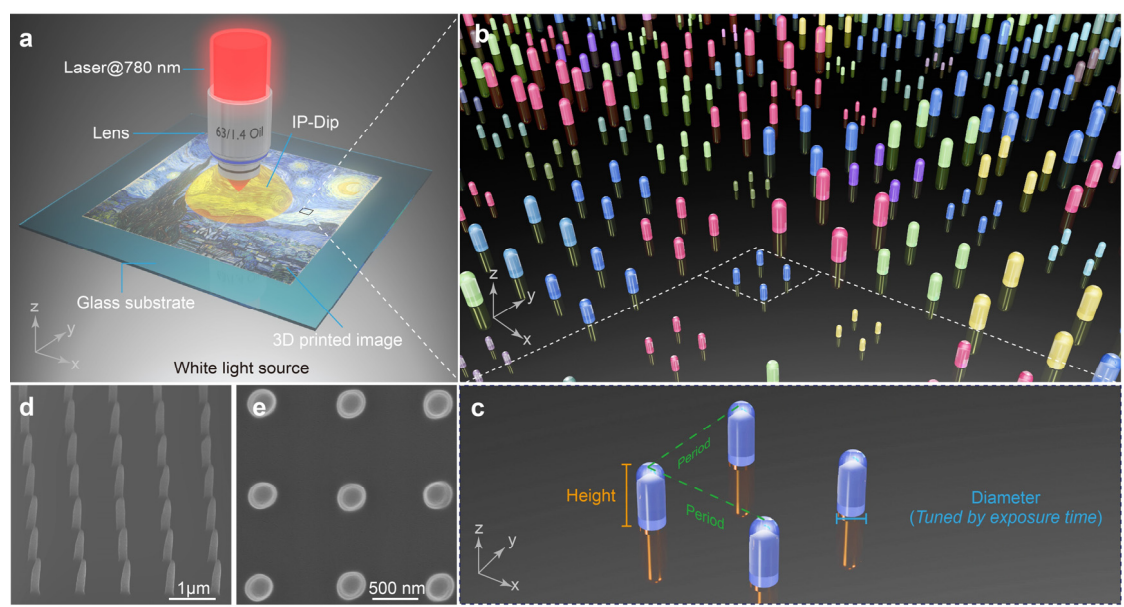

**Fig. 1 Nanoscale full colour and grayscale 3D printing. a** Schematic of the fabrication process using the Photonic Professional GT system (Nanoscribe GmbH). A collimated femtosecond laser at wavelength of 780 nm is focused through an immersion objective lens (63×, NA = 1.4) to induce two-photon polymerization in photoresist IP-Dip. After development, arbitrary microscopic colourful prints are observed under backlit white light illumination. **b, c** The printed image is formed by nanopillars designed with different heights, periodicities, and diameters. **d, e** Scanning electron micrographs of printed nanopillar array at 45° tilted view and top view, respectively.

**Results and discussion**

Nanoscale 3D printing by TPL was achieved using the Photonic Professional GT system (Nanoscribe GmbH). As shown in the schematic in Fig. 1a, a femtosecond pulsed laser at wavelength of 780 nm is focused down to a diffraction-limited spot by an immersion objective lens (63×, numerical aperture (NA) = 1.4), inducing crosslinking at the focal point within the liquid resin IP-Dip. The liquid resin fills the space between the glass substrate and the lens, in the "dip-in" configuration. To achieve the highest print resolutions, we used the pulsed mode to expose a single voxel (volumetric pixel) at a time. Figure 1b shows part of the schematic of desired result consisting of nanopillars of different sizes, heights and periods achieved by 3D printing one segment and one layer at a time.

To quantify the effect of nanopillar size on the colours, the HTP parameters were systematically varied in a single fabrication run (Fig. 1c). The nanopillar diameters were tuned by varying the point exposure duration of the laser beam. The samples were rinsed after exposure (see Methods) and the printed images consisting of nanopillars were inspected under a brightfield optical microscope in transmission mode illumination. Scanning electron micrographs (SEM) of a representative nanopillar array are shown in Fig. 1d and e, demonstrating that nanoscale pillars can be printed as design using TPL.

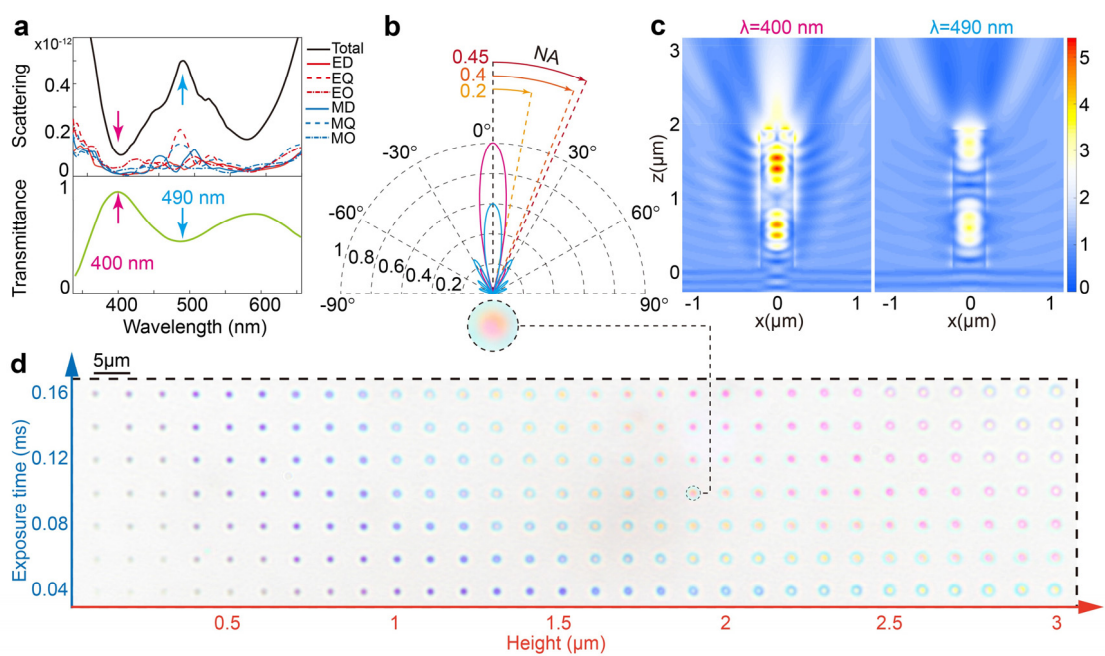

**Fig. 2 Scattering and transmittance spectra, and colours from single nanopillars. a** Simulated scattering and transmittance spectra with multipole decomposition analysis corresponding to that of a nanopillar printed with diameter 420 nm (exposure time 0.10 ms at laser power of 25 mW) and height

1.9 μm. The integration of energy in the far field is performed over a solid angle equivalent to a NA of 0.4 to match the 50× objective lens used. **b** Normalized far field angular projections of the transmitted light intensity at wavelengths of 400 nm (magenta) and 490 nm (blue). The dashed lines represent the collection half-angle of different objective lenses: 10×, NA = 0.2 (light orange dashed line); 50×, NA = 0.4 (orange dashed line); 20×, NA = 0.45 (red dashed line). **c** Electric field map of light incident on a single nanopillar at wavelengths of 400 nm (left panel) and 490 nm (right panel). **d** Optical micrograph of colours from an array of single nanopillars with heights varying from 0.1 μm to 3 μm (steps of 0.1 μm), and exposure time ranging from 0.04 ms to 0.16 ms in steps of 0.02 ms, fabricated at laser power of 25 mW. Sample was backlit and imaged using a 50×, NA = 0.4 objective lens. The magnified image (marked with black dashed line) shows the colour distribution corresponding to Fig. 2b.

To analyze the optical response of the nanopillar, we performed transmittance and scattering simulations with multipolar decomposition[45] of a single nanopillar with diameter 420 nm (exposure time 0.10 ms at laser power of 25 mW) and height 1.9 μm, as shown in Fig. 2a. The mechanism for colour generation from a single nanopillar has yet to be studied and is more complex than the simple diffractive effects previously claimed to be the dominant effect at play[46]. Here, we show that scattering is the dominant contributor to the spectral response. Multipolar decomposition shows the scattering contributions from the electric dipole (ED), magnetic dipole (MD), electric quadrupole (EQ), magnetic quadrupole (MQ), electric octupole (EO), magnetic octupole (MO), and other higher order modes. The representative local dip (400 nm) and peak (490 nm) in the simulated total scattering cross section (top panel of Fig. 2a) align with the peak and valley in the simulated transmittance spectrum respectively (bottom panel of Fig. 2a). Furthermore, the peak and valley in the transmittance spectrum also could be confirmed by the far field projection of the transmitted light in Fig. 2b. For a wavelength of 400 nm, most of the intensity is confined within a narrow forward lobe (solid magenta line) that fits within the collection cone of the objective lens (10×, NA = 0.2 light orange dashed line; 50×, NA = 0.4 orange dashed line; 20×, NA = 0.45 red dashed line), and thus a strong peak appears in the transmittance spectrum and almost remains unchanged with increasing NA. Conversely, a large proportion of the incident light is strongly scattered outside of the collection cone of the objective lens (blue solid line) at 490 nm, resulting in a dip in the transmittance spectrum. As only the main forward lobe is collected for NA < 0.45, the spectra and colours obtained by the aforementioned objective lenses are similar. Figure 2c shows the corresponding electric field magnitudes (normalized to incident electric field) of the single nanopillar at wavelengths of 400 nm and 490 nm, consistent with results in Figs. 2a and 2b. Figure 2d shows the transmissive colours (captured by 50×, NA = 0.4 objective lens) of an array of single nanopillars with nominal height varying from 0.1 μm to 3 μm in steps of 0.1 μm along the *x* axis, and exposure time from 0.04 ms to 0.16 ms in steps of 0.02 ms along the *y* axis. The magnified image (marked with black dashed line) shows the captured colour corresponding to Fig. 2a, in which the colour changes from light magenta (centre) to light orange (edge), in accordance with the angular intensity distribution in the collection cone in Fig. 2b. With the optimized process parameters, nanopillar aspect ratios as high as ~9 can be achieved. The primary colour of single nanopillars varies dramatically from light grey to blue, orange, magenta, and green with the increase in height, and changes more slowly with increasing exposure time (diameter), demonstrating that colors are generated from single nanopillars.

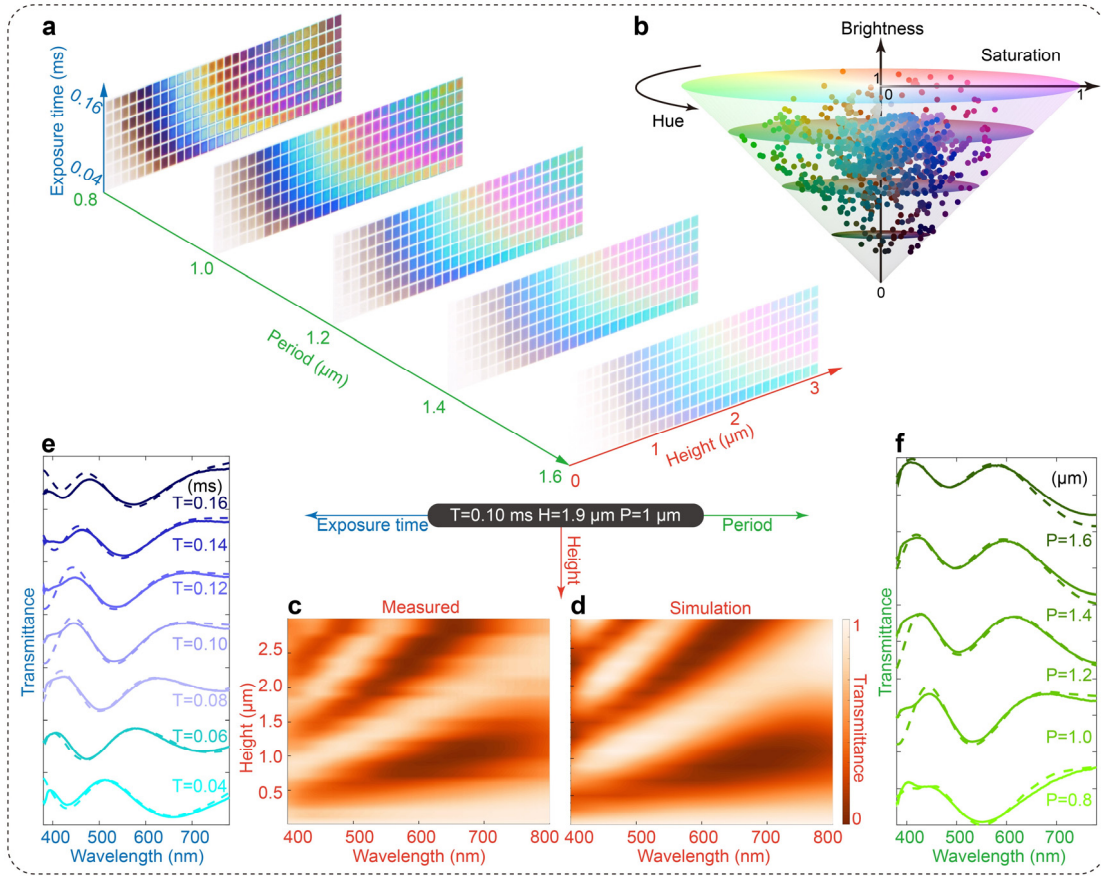

**Fig. 3 3D full colour palette and spectra from arrays of nanopillars. a** 3D transmissive colour palette obtained by tuning the height, period, and exposure time of nanopillars in arrays. The laser power used was 25 mW and the size of each colour patch is 20 μm², captured by a 10×, NA = 0.2 objective lens. **b** Measured spectra from the colour patches in (a) plotted in and occupying most of the 3D HSB colour space. **c** Measured and **d** simulated transmittance spectra of nanopillar arrays with fixed exposure time 0.10 ms, period 1 μm, and height ranging from 0.1 μm to 3 μm in steps of 0.1 μm. **e** Simulated (dashed lines) and measured (solid lines) transmittance spectra of nanopillar arrays with designed height 1.9 μm, period 1 μm, and exposure time ranging from 0.04 ms to 0.16 ms in steps of 0.02 ms. **f** Simulated (dashed lines) and measured (solid lines) transmittance spectra of nanopillar arrays with designed height 1.9 μm, exposure time 0.10 ms, and period ranging from 0.8 μm to 1.6 μm in steps of 0.2 μm.

Based on the analysis above and the results in Fig. 2d, we experimentally obtained a 3D full colour palette by gradually tuning the third parameter of the nanopillar array, i.e., period, from 0.8 μm to 1.6 μm in steps of 0.2 μm, as shown in Fig. 3a. With increasing period, the colours become less saturated. This result consistently shows that individual pillars act as colour generators, such that spacing them further apart has the effect of "diluting" the colour, with negligible change in their hues. By mapping the 1,050 measured spectra from the HTP based 3D full colour palette in Fig. 3a into the HSB colour space (Fig. 3b), colours with different HSB values are obtained, including very dark and saturated ones (see also CIE1931 chromaticity diagram in Supplementary Fig. S1, and a 3D rotating animation of Fig. 3b's HSB colour space plot in Supplementary Movie S1). The results indicate that a single printing process of this simple design generates an abundance of colours and grayscale values to achieve a wide coverage of the full HSB colour space.

To fully understand the influence of the HTP parameters, we independently tuned them starting from the nanopillar array with the parameter set of exposure time 0.10 ms, height 1.9 μm, and period 1.0 μm (see spectra comparison with silicon and silver nanopillars in the same geometry in Fig. S2). Firstly, we measured the transmittance spectra of nanopillars as a function of height as shown in Fig. 3c, showing excellent agreement with simulations in Fig. 3d. It is noteworthy that nanopillars shorter than ~ 0.7 μm exhibit spectra that are almost flat due to the weak scattering of the nanopillar in the visible band. As the height increases to 0.7 μm, the transmittance decreases, resulting in the multiple shades of grey (Fig. 3a). Beyond this range, resonance peaks appear and redshift with taller nanopillars, with resulting hues varying from blue to cyan, orange, and magenta. With higher order resonances excited inside taller pillars, we also observe pink, green, and brown (Fig. 3a). Secondly, we varied the exposure time as shown in Fig. 3e. With the increase of exposure time, the nanopillar diameter increases logarithmically and eventually flattens out (see measured diameters and fitting model in Supplementary Fig. S3)[47]. This trend corresponds to the larger redshifts in the spectra for exposure times < 0.10 ms in Fig. 3e compared to spectral shifts for exposure time > 0.10 ms (varying exposure time for a constant nominal height is accompanied by slight height variations as discussed in Supplementary Note S1 and Supplementary Fig. S4). This effect is also confirmed by the colour evolution in Fig. 3a. Lastly, with the increasing periodicity, the areal coverage of nanopillars is reduced, thus more of the incident light is simply transmitted without scattering, i.e., the nanopillar array has a smaller effective scattering cross section, reducing the spectral contrast (Fig. 3f). Deviations between the simulated and measured spectra in Fig. 3c–f are likely due to fabrication imperfections, such as altered morphology of nanopillars induced by laser fluctuation, inhomogeneous response of the photoresist, and mechanical vibrations.

Based on the full colour palette in Fig. 3a, a parameter matching algorithm was developed for 3D colour printing, as illustrated in Fig. 4a. Reminiscent of digital image processing, in which an image is split into RGB channels, here we map an image into height, exposure time, and period, i.e., three channels of HTP for 3D printing. The HTP values of each pixel were retrieved by finding the shortest distance between the target colour and the full colour palette in the $L^*a^*b^*$ colour space, which is suited for determining colour differences (see the sampled colour palette, original images, and colour retrieving comparisons with different colour spaces in Supplementary Figs. S5–S7). Then a script was written to convert HTP values into the instruction set for Nanoscribe to control the scanning of femtosecond pulsed laser and movement of the piezoelectric stage. To make full use of the available colours, the pixel size was set to 3.2 μm. The optical micrograph of a backlit sample in Fig. 4b shows the printed colourful painting of Vincent van Gogh's Starry Night (10× objective lens). The SEM images captured by top view in Fig. 4c and 45° tilted view in Fig. 4d of the same section clearly show pillars of varying HTP values.

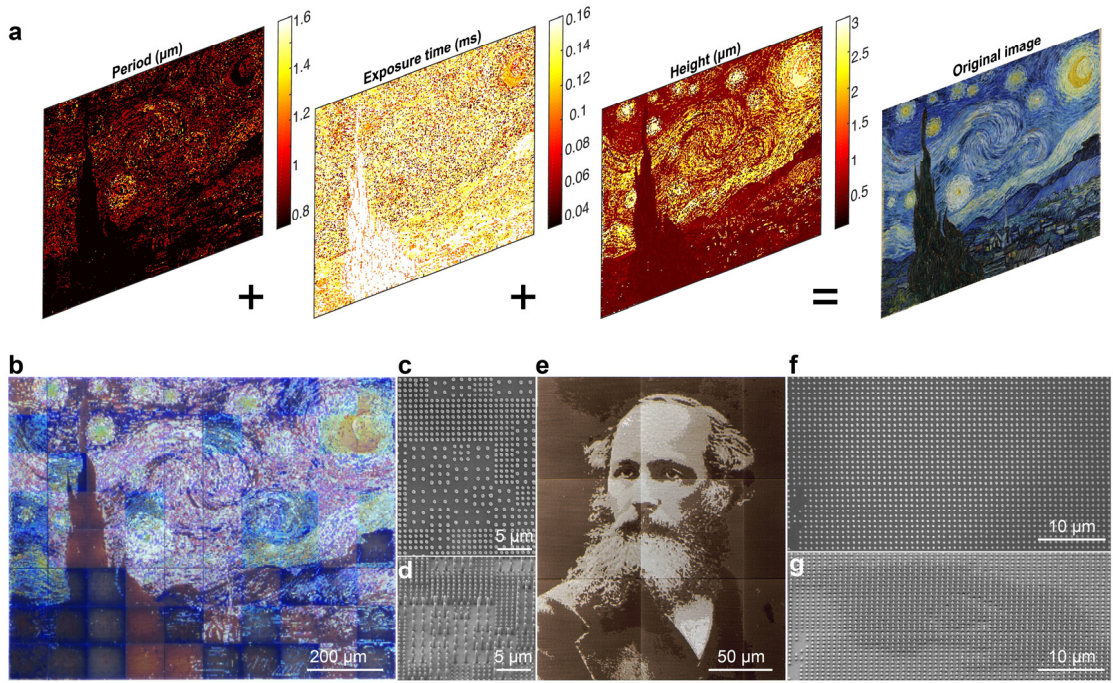

**Fig. 4 Colour matching strategy for colourful and grayscale 3D printing. a** Pixel-by-pixel mapping of an image into height, exposure time, and period channels selected by colour matching to the experimental colour palette for each pixel. **b** Transmission optical micrograph of 3D printed colourful painting of the Starry Night (Vincent van Gogh, 1889, 300×240 pixels, 10× objective lens). **c** Top view and **d** 45° tilted view scanning electron micrographs of the same section in (b), showing pillars of varying heights, diameters and periodicities. **e** Transmission optical micrograph of grayscale portrait of James Clerk Maxwell (240×300 pixels, 10× objective lens). **f** Top view and **g** 45° tilted view scanning electron micrographs of the structures that form his left eye in (e), showing pillars of varying heights and diameters.

Figure 4e shows the painted grayscale portrait of James Clerk Maxwell with nanopillars at fixed period of 0.8 μm, while heights and exposure time were freely varied, as shown from the SEM images in Figs. 4f and 4g (also see the full SEM image in Supplementary Fig. S8). Each pixel is represented by a single nanopillar and 18 gray levels are achieved in the print. The details of original images are preserved despite the tiled appearance, which is caused by variations across different write fields (100 × 100 μm$^2$ for Fig. 4b, 80 × 80 μm$^2$ for Fig. 4e, customizable according to the design) that led to stitching errors from the movement of the motorized stage. Deviations in colours in adjacent write fields are a limitation of the equipment, caused by the inaccuracy of determining the interface between the substrate and the resin that leads to systematic height variations (see fabrication details in Methods and Note S2 in Supplementary Information)[48]. The experimental results confirm the effectiveness of controlling HSB through HTP parameters, achieving with high quality 3D printed nanopillars as laid out in the design strategy of Fig. 1b. In addition, the full colour and grayscale 3D prints are readily reproducible with our methods, as also demonstrated in Supplementary Fig. S9.

The understanding that the nanopillars act as directional scatterers makes it possible to realize different visual effects and steganography with the use of different illumination conditions. The two generally used brightfield and darkfield illumination configurations in optical microscopes are depicted in Fig. 5a. We 3D printed the famous Chinese calligraphy artwork *Preface to the Poems Composed at the Orchid Pavilion* (Lantingji Xu, Shenlong version) by Wang Xizhi (traditionally referred to as the Sage of Chinese Calligraphy) to show the illumination-based visual effects. The period was fixed at 0.8 μm, the exposure time was set as 0.10 ms, and each pixel was represented by a single nanopillar with dimensions that produces the closest colour to the desired pixel. The sample was imaged using a 20× objective lens with brightfield transmission mode illumination (Fig. 5b) and darkfield reflection mode illumination (Fig. 5c). These illumination modes reproduce the two primary artistic forms of Chinese calligraphy (see SEM image in Fig. 5d): cocoon paper calligraphy (white background) and stone rubbing (black background) (the print is around 1/465 scale of the original versions, see Supplementary Figs. S10, S11, printing process in Supplementary Movie S2). As seen from the brightfield image, the details of grayscale Chinese ink and the red seals are preserved in the miniaturized reproduction. In darkfield, we observe grayscale inversion where the background (characters) switches from white (black) to black (white), and the red of the seal is desaturated to light gray. The bright appearance of the characters is caused by the broadband scattering that only occurs at the nanopillars, and the scattering spectra are similar for nanopillars with the chosen heights (see scattering spectra in Supplementary Fig. S12).

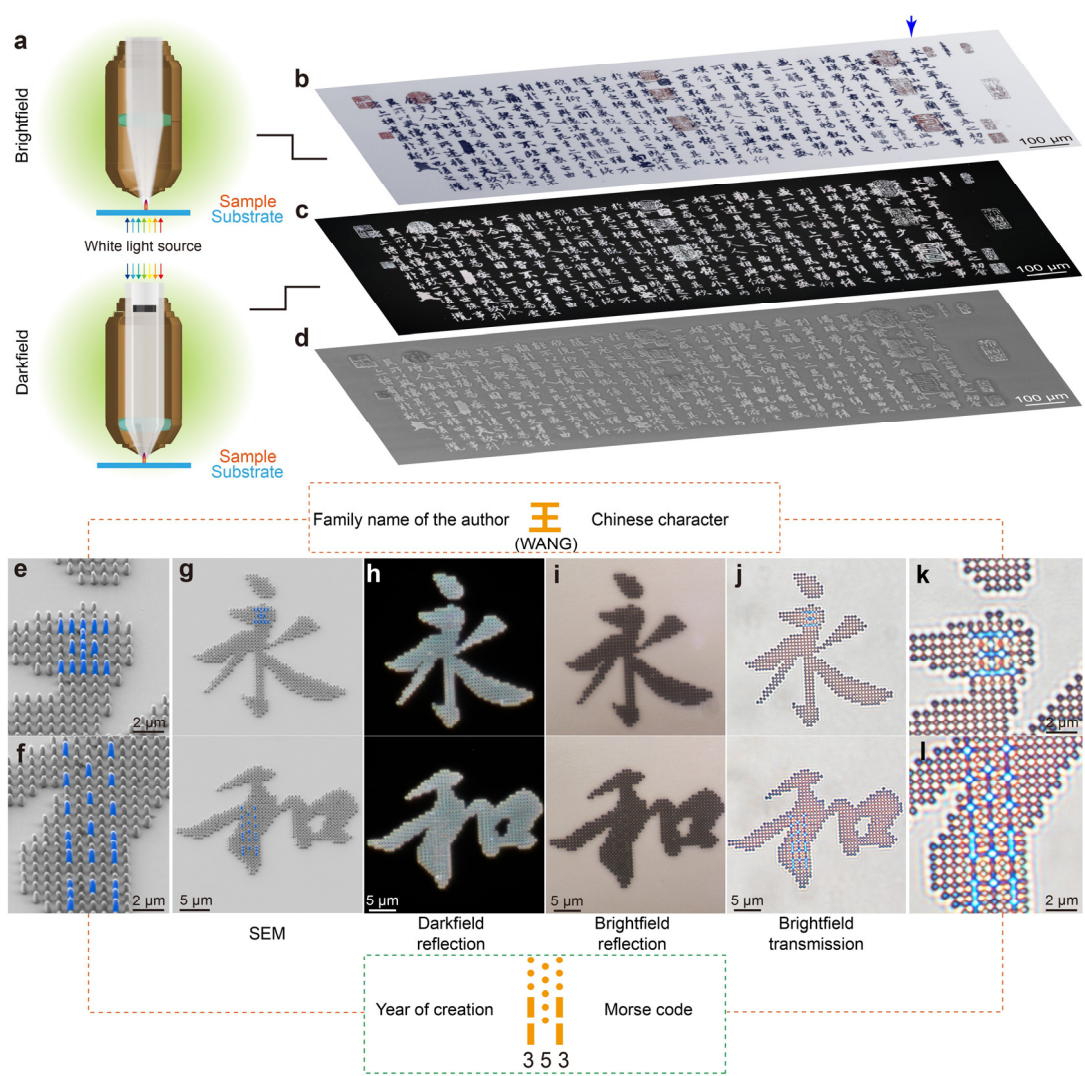

**Fig. 5 Steganography in 3D printed brightfield and darkfield images of Chinese calligraphy. a** Brightfield and darkfield illumination configurations. **b** Brightfield transmission and **c** darkfield reflection micrographs of the printed Chinese calligraphy *Preface to the Poems Composed at the Orchid Pavilion* (Lantingji Xu) by Wang Xizhi (2000×658 pixels), captured by a 20× objective lens. The two images are microscale representations of the two primary artistic forms of Chinese calligraphy, i.e.,

cocoon paper calligraphy (white background) and stone rubbing (black background). The final images shown here are obtained by stitching together multiple micrographs. Printing parameters: period 0.8 μm and exposure time 0.10 ms. **d** scanning electron micrograph of the printed Chinese calligraphy. **e** The family name of the author in Chinese character "王" and **f** the year of creation 353 in Morse code have been secretly encoded as hidden messages into the first two Chinese characters in the top-right corner of the calligraphy (position marked with a blue arrow) in **b**, **c**, **d**. **g** SEM images with false coloured nanopillars (in blue) representing the hidden information. **h** Darkfield and **i** brightfield images via reflection mode illumination of the two Chinese characters by a 100× objective lens. **j** Brightfield image observed by a 100× objective lens with transmission mode illumination, revealing the hidden information in the taller pillars that appear blue. **k, l** Magnified view of (j). Designed heights of nanopillars in (e–j): 0.9 μm for hidden information, and 0.5 μm for others.

As a single nanopillar can be used to generate colour (Fig. 2d), we increased the height of a few selected nanopillars from 0.5 μm to 0.9 μm to conceal additional information inside the first two Chinese characters "永和" (which means "forever harmony"), indicated by a blue arrow in Fig. 5b. The heights were carefully chosen to effectively hide the information and only reveal it clearly in brightfield transmission mode. Magnified SEM images of the Chinese characters are shown in Figs. 5e–g. In the first Chinese character "永", the family name of the author "王" (WANG) was encoded with 15 nanopillars at the height of 0.9 μm. In the second Chinese character "和", the year of creation 353 was concealed in Morse code (single nanopillar represents one dot and three neighboring nanopillars represent one dash). As seen from the optical micrograph captured by a 100× objective lens with darkfield illumination (Fig. 5h), the information is concealed as the scattering spectra appear nearly identical (see scattering spectra in Supplementary Fig. S12). The information is also hidden in brightfield reflection mode illumination as shown in Fig. 5i, because most of the incident light is either transmitted or scattered (see reflectance spectra in Supplementary Fig. S12). Only in brightfield transmission mode illumination can we reveal the secret messages, as presented in Fig. 5j and magnified in Figs. 5k and 5l. Because the nanopillars generate colours independently, the taller nanopillars appear blue and can be distinguished from the surrounding shorter ones. These colors are consistent with the colors of nanopillars of the same heights in Figs. 2d and 3a. By careful design of the HTP values for each nanopillar, we could hide even more complex information and multilevel messages within the prints. The results suggest that the height-based out-of-plane encoding provides a reliable new method for anti-counterfeiting, with increased secrecy beyond the traditional in-plane design. The grayscale inversion between brightfield and darkfield illuminations and colour desaturation in darkfield could find applications in steganography for unobtrusive protection of valuable items. Notably, the prints have a near colorless appearance when observed by eye under ambient illumination because of strong scattering (Supplementary Fig. S13), thus providing further secrecy to the prints.

**Conclusion**

TPL based 3D printing provides access to a full 3D colour palette with control of hue, saturation, brightness, and grayscale shades by precise tuning of the height, diameter, and spacing of nanopillar arrays. The color control down to the single nanopillar preserves the subwavelength footprint of colour-generating low-refractive-index dielectrics. Despite the low-refractive-index, these nanostructures are effective scatterers with spectrally selective directivity. Information encoded in submicron variations in height of the nanopillars are concealed in brightfield and darkfield reflection but are revealed only in brightfield transmission microscopy, thus enabling steganography. We expect that 3D printed nanostructures more elaborate than the nanopillar would potentially exhibit polarization and chiral effects, in addition to enhanced colour gamut. These functions can be readily extended to the infrared regime with promising applications in on-chip optoelectronics, integrated photonic circuity, anti-counterfeiting, and sensing.

**Methods**

**Fabrication of structures with two-photon polymerization lithography.** The 3D printer Nanoscribe GmbH (Photonic Professional GT system, Nanoscribe Inc., Germany) was used to print the nanopillar structures. An immerse objective lens (Zeiss Plan Apo 63×, NA = 1.4) was used to focus femtosecond laser pulses (pulse duration 100 fs, frequency 80 MHz, laser power 50 mW, and wavelength of 780 nm) into the negative photoresist IP-Dip (Nanoscribe Inc., Germany) on a fused silica glass substrate (25 × 25 $mm^2$ and thickness of 0.7 mm, Haian Huihong, China) to induce two-photon polymerization. The beam position in the horizontal $xy$ plane within one write field (80/100 μm in our printing, can be customized) was tuned by a group of galvanometric mirrors and a piezoelectric translation stage was used to control the $z$ position. The movement between different write fields was conducted by a motorized stage. An optimal laser power of 25 mW (50% of the total power) was used during prints. After exposure, the samples were developed in polyethylene glycol methyl ether acetate for 10 min to remove the unexposed photoresist, then immersed in isopropyl alcohol for 5 min, lastly transferred into nonafluorobutyl methyl ether for 5 min to reduce the effect of surface tension during the evaporation of liquid. The samples were finally dried in the air for several seconds. The three chemicals used in post-development were from Sigma-Aldrich. The time taken for colourful painting of Starry Night, grayscale portrait of James Clerk Maxwell, and Chinese calligraphy were 2 h 16 min, 12 min, and 27.5 min, respectively.

**Code.** A Matlab script was written to convert height, exposure time, and period values into the instruction set in gwl format for Nanoscribe to control the scanning of femtosecond pulsed laser and movement of the piezoelectric stage. For printing with unfixed period, the pixel size was set to be 3.2 μm, and the number of pillars in each direction was rounded to the nearest integer, i.e., round(pixel size/period). The position of the first nanopillar was set as (pixel size - period(pixel number - 1))/2 to reduce the color difference.

**Spectra measurement.** The transmittance, reflectance, and scattering spectra, along with optical micrographs were taken via a Nikon Eclipse LV100ND optical microscope equipped with a Nikon DS-Ri2 camera and a CRAIC 508 PV microspectrophotometer. Two halogen lamps (LV-HL 50 W) were used to illuminate the samples in reflection and transmission modes, respectively. The transmittance spectra were normalized with a bare fused silica substrate, and the reflectance spectra were normalized with an aluminium mirror. The specifications of objective lens used for measurement were: 10×, NA = 0.2; 20×, NA = 0.45; 50×, NA = 0.4; and 100×, NA = 0.9.

**Characterization.** Scanning electron microscopy was performed with JEOL-JSM-7600F and Raith eLine Plus systems.

**Simulation.** The spectra simulation was conducted with a commercial Finite-Difference Time-Domain software (Lumerical, Canada). IP-Dip and fused silica substrate were modeled with the experimentally measured dispersive refractive index[49]. To be consistent with the experiment, non-periodic boundary conditions were used with a total-field scattered-field source normally incident from the substrate side. The nanopillars were modelled as cylinders with corresponding height, diameter, and period. A field and power monitor was placed right upon the structures in simulation to record the electromagnetic field. Afterwards, near-to-far-field projection was performed to integrate the transmitted light within the collection cone of the experimentally used objective lens. Normalization was performed in the same process as the experiment. Multipole decomposition analysis was carried out using a Finite Element Method based software (COMSOL Multiphysics 5.4, Sweden). The scattering and field distributions were double checked with both software.

**Data availability**

All data are available from the corresponding author upon reasonable request.

**Acknowledgements**

This research is supported by National Research Foundation (NRF) Singapore, under its Competitive Research Programme NRF-CRP001-021 and CRP20-2017-0004, Singapore University of Technology and Design (SUTD) Digital Manufacturing and Design (DManD) Center (RGDM 1830303), and THALES-SUTD projects RGTHALES1801 and RGTHALES1901.

**Author Contributions**

H.W. conceived the idea, developed the algorithm and generated the 3D printing files, performed the design, fabrication, measurement and characterization. Q.F.R., H.T.W., H.L.L., W.Z., J.T., and J.Y.E.C. assisted the fabrication and characterization. Q.F.R. and H.T.W. helped the spectra measurement. H.W., Q.F.R., and K.T.P.L. conducted the FDTD simulations, and S.D.R. did the multipole decomposition calculation. J.K.W.Y. supervised the research. All the authors contributed to the writing of the manuscript.

**Conflict of Interest**

The authors declare no conflict of interest.